\documentclass[12pt,eqsecnum]{revtex4}
\begin{document}
\title{Relating Gribov-Zwanziger theory to effective Yang-Mills theory}

\author{Sudhaker Upadhyay\footnote {e-mail address: sudhakerupadhyay@gmail.com}} 
\author{Bhabani Prasad Mandal\footnote{e-mail address: bhabani.mandal@gmail.com }}

\affiliation { Department of Physics,\\
Banaras Hindu University,\\
Varanasi-221005, INDIA. \\
}

\begin{abstract}
We consider the Gribov-Zwanziger (GZ) theory with appropriate horizon term which exhibits 
the nilpotent BRST invariance. This infinitesimal BRST transformation has been 
generalized by allowing the parameter to be finite and field dependent (FFBRST).
By constructing appropriate finite field dependent parameter we show that the generating
functional of GZ theory with horizon term is related to that of Yang-Mills (YM) theory 
through FFBRST transformation.
\end{abstract}
\maketitle

\section{Introduction}
In order to quantize a gauge theory it is necessary to eliminate the redundant degrees of 
freedom from the functional integral representation of the generating functional. This 
can be done by modifying the generating functional with the addition of a gauge fixing term 
\cite
{fp}.
However in non-Abelian gauge theories even after gauge fixing the redundancy of gauge 
fields is not completely removed in certain gauges for large gauge fields (Gribov problem)
\cite{gri}. The non-Abelian gauge theories in those gauges contain so-called Gribov copies. 
Gribov copies play 
a crucial role in the infrared (IR) regime while it can be neglected in the perturbative 
ultraviolet (UV) regime \cite{gri, zwan, zwan2}.
This topic has become very exciting currently due to the fact that color confinement is closely 
related to the asymptotic behaviour of the ghost and gluon propagators in deep IR regime
\cite{kon0}.

In order to resolve the Gribov problem, Gribov and Zwanziger  proposed a 
theory, which restricts the domain of integration in the functional integral within
 the first Gribov horizon \cite{zwan}. The restriction to the  Gribov region $\Omega $ can be
achieved by adding a nonlocal term, commonly known as horizon term, to the YM action \cite
{zwan, zwan1, zwan2}.
 
The Kugo-Ojima (KO) criterion for color confinement \cite{ko} is based on the assumption of an 
exact BRST invariance of YM theory in the manifestly covariant gauge.
But the YM action restricted
in Gribov region (i.e. GZ action) does not exhibit the usual BRST invariance, due to the
presence of the nonlocal horizon term \cite{sore1}. 
Recently, a nilpotent BRST transformation which leaves the GZ 
action invariant has been obtained and  
can be applied to KO analysis of the GZ theory \cite{sor}.
The BRST symmetry in presence of the Gribov horizon has great applicability in order to 
solve the nonperturbative features of confining YM theories \cite{dud, fuj},
where the soft breaking of the BRST symmetry exhibited by the GZ action can be converted 
into an exact invariance \cite{sor1}. Such a modification is very useful in order to
evaluate the vacuum expectation value (VEV) of BRST exact quantity.

In this work we generalize the nilpotent BRST transformation introduced in Ref. \cite{sor}
for GZ theory by allowing the parameter to be finite field dependent following the method 
developed by 
Joglekar and Mandal for pure YM theory
 \cite{sdj}. Such a generalized BRST (FFBRST) transformation is nilpotent and leaves the 
 effective action invariant. However, being finite in nature such a transformation 
 does not leave the path integral measure and hence the generating functional invariant.
By constructing an appropriate finite field dependent parameter we show that such 
FFBRST transformation relates the generating functional for GZ theory to the generating 
functional in YM theory. 
 
The paper is organized in the following manner. In Sec. II we illustrate some of the 
essential features in GZ theory. In Sec. III we discuss  the nilpotent 
BRST transformation of the multiplicative renormalizable GZ theory. Sec. IV is devoted to 
the discussion of finite field dependent 
BRST transformation in Euclidean space. Connection of GZ theory and YM theory is 
established in Sec. V.
Last section contains discussions and conclusions.

\section{GZ theory: brief introduction}
It has been shown in Ref. \cite{zwan2} that the restriction to the Gribov region $\Omega$ (defined in such a way 
that
the Faddeev-Popov (FP) operator is strictly positive. i.e.
 \begin{equation}
\Omega \equiv \{ A_\mu^a, \partial_\mu A_{\mu}^ a=0, {\cal{M}}^{ab} >0\} ),
 \end{equation} 
can be imposed by adding a nonlocal term $S_h$, given in Eq. (\ref{gact}) below,
 to the standard YM 
action
\begin{equation}
S_{YM}=S_0+S_{GF+FP},
\end{equation}
where $S_0$ is the kinetic part and $S_{GF+FP}$ is the ghost and gauge (Landau gauge) fixing 
part of the YM action respectively,
\begin{eqnarray}
S_0&=&\int d^4x\left[\frac{1}{4}F^a_{\mu \nu }F_{\mu \nu}^a\right],\nonumber\\
S_{GF+FP}&=&\int d^4x \left[B^a\partial_\mu A_\mu ^a +\bar c^a\partial_\mu {\cal D}_\mu 
^{ab}c^b\right].
\end{eqnarray}
The nonlocal term in $4$-dimensional Euclidean space is written as
\begin{equation}
S_h=\int d^4x h(x), \label{gact}
\end{equation}
where the integrand $h(x)$ is the horizon function. There exist different choices
for the horizon function  in literature \cite{sor}. One such horizon term is
\begin{equation}
h_1(x)=\gamma^4\int d^4y\ g^2 f^{abc}A_\mu^b(x) ({\cal{M}}^{-1})^{ce}(x,y) f^{ade}A^{d}_\mu 
(y).
\label{ht1}
\end{equation}
$({\cal{M}}^{-1})^{ce}$ is the inverse of the Faddeev-Popov operator 
${\cal{M}}^{ab}\equiv -\partial_\mu {\cal D}^{ab}_\mu  =-\partial_\mu 
(\partial_\mu\delta^{ab}+gf^{acb}A^c_\mu)$. The Gribov parameter $\gamma$ can be obtained in 
a consistent way by solving a gap equation (also known as horizon condition)
\begin{equation}
\left<{h(x)}\right> = 4(N^2-1), \label{hc}
\end{equation}
where $N$ is the number of colors.
Another horizon term which gives the correct multiplicative
renormalizability of the GZ theory is given as \cite{kon1, sor}
\begin{eqnarray}
h_2(x)&=&\lim_{\gamma(x)\rightarrow \gamma}\int d^4y \left[\left({\cal D}_\mu^{ac}(x)
\gamma^2(x)\right) 
({\cal{M}}^{-1})^{ce} (x,y)\left({\cal D}_\mu^{ ae}(y)\gamma^2(x)\right)\right].\label{ht2}
\end{eqnarray}
The nonlocal term (\ref{gact}) corresponding to the horizon function (\ref{ht2}) can be 
localized as 
\cite{zwan, zwan2}
\begin{equation}
e^{-S_{h_2}}=\int {\cal D}\varphi{\cal D}\bar\varphi{\cal D}\omega{\cal D}\bar\omega
e^{S_{loc}},
\end{equation}
with
\begin{eqnarray}
S_{loc}&=&\int d^4x\left[\bar\varphi_i^{a}\partial_\mu{\cal D}_\mu^{ab}
\varphi^{b}_i-\bar\omega_i^{a}\partial_\mu{\cal D}_\mu^{ab}
\omega^{b}_i-\gamma^2{\cal D}_\mu^{ca} (\varphi_\mu^{ac}(x)\bar\varphi_\mu^{ac}(x))\right],
\end{eqnarray}
where a pair of complex conjugate bosonic field  $ 
(\bar\varphi_i^{a}, \varphi_i^{a})=(\bar\varphi_\nu^{ac}, \varphi_\nu^{ac})$ 
and anticommuting auxiliary fields
$( \omega_i^{a}, \bar\omega_i^{a})=( \omega_\nu^{ac}, \bar\omega_\nu^{ac})$, with composite 
index $i=(\nu, c)$, have been introduced. As at the level of the 
action, total derivatives are always neglected, $S_{loc}$ becomes
\begin{eqnarray}
S_{loc}&=&\int d^4x\left[\bar\varphi_i^{a}\partial_\mu{\cal D}_\mu^{ab}
\varphi^{b}_i-\bar\omega_i^{a}\partial_\mu{\cal D}_\mu^{ab}
\omega^{b}_i\right.\nonumber\\
 &-&\left.\gamma^2gf^{abc}A_\mu^a (\varphi_\mu^{bc}(x) +\bar\varphi_\mu^{bc}(x))\right].
\end{eqnarray}
Here it is concluded that at the local level horizon functions (\ref{ht1}) and (\ref{ht2}) 
are same.
So that the localized GZ action becomes
\begin{eqnarray}
S_{GZ}&=& S_{YM} +S_{loc}\nonumber\\
&=& S_{YM} +\int d^4x\left[\bar\varphi_i^{a}\partial_\mu{\cal D}_\mu^{ab}
\varphi^{b}_i-\bar\omega_i^{a}\partial_\mu{\cal D}_\mu^{ab}
\omega^{b}_i\right.\nonumber\\
&-&\left.\gamma^2gf^{abc}A_\mu^a (\varphi_\mu^{bc} +\bar\varphi_\mu^{bc})\right].
\end{eqnarray}
 Thus the local action $S_{GZ}$ and the nonlocal action $S_{YM}+S_h$ are related as
 the following 
 \begin{equation}
\int [D\phi_1]e^{-\{S_{YM}+S_{h_2}\}}=\int [D\phi]e^{-S_{GZ}},\label{nlact}
\end{equation}
with $\int[D\phi_1]\equiv \int[DA DB Dc D\bar c]$ and $\int[D\phi]\equiv \int[DA DB Dc D\bar cD\varphi D\bar\varphi
D\omega D\bar\omega]$.
By differentiating Eq. (\ref{nlact}) with respect to $\gamma^2$ and noting $\left < 
\partial_\mu \varphi^{aa}_\mu \right > = \left < \partial_\mu \bar\varphi^{aa}
_\mu \right >=0$, 
 the horizon condition in Eq. (\ref{hc}) is recast as 
\begin{equation}
\left<gf^{abc}A_\mu^a (\varphi_\mu^{bc}+\bar\varphi_\mu^{ bc})\right> +8\gamma^2(N^2-1)
  =0.
\end{equation}
The horizon condition can further be written as \cite{sor,sor1}
 \begin{equation}
\frac{\partial \Gamma }{\partial \gamma^2}=0,\label{hori}
\end{equation}
with $\Gamma $, the quantum action defined as 
\begin{equation}
e^{-\Gamma }=\int [D\phi] e^{-S_{GZ}}.
\end{equation}

We see that the horizon condition (\ref{hori}) is equivalent to
\begin{eqnarray}
\left < 0\mid gf^{abc}A_\mu^a\varphi^{bc}_\mu \mid 0\right > &+&\left < 0\mid gf^{abc}
A_\mu^a\bar\varphi^{ bc}_\mu\mid 0\right >\nonumber\\
 &=&-8\gamma^2(N^2-1),
\end{eqnarray}
which, owing to the discrete symmetry of the action $S_{GZ}$ 
\begin{eqnarray}
\bar\varphi^{ac}_\mu&\rightarrow & \varphi^{ac}_\mu\nonumber\\
\varphi^{ac}_\mu&\rightarrow & \bar\varphi^{ac}_\mu\nonumber\\
B^a&\rightarrow & (B^a -gf^{amn}\bar\varphi^{mc}_\mu \varphi^{nc}_\mu ),
\end{eqnarray}
becomes
\begin{eqnarray}
\left < 0\mid gf^{abc}A_\mu^a\varphi^{bc}_\mu \mid 0\right > &=&\left < 0\mid gf^{abc}
A_\mu^a\bar\varphi^{ bc}_\mu\mid 0\right >\nonumber\\
 &=&-4\gamma^2 (N^2-1).\label{exp}
\end{eqnarray}
Further the constant term $4\gamma^4(N^2-1)$ is introduced in $ S_{GZ}$, to incorporate 
the effect of horizon condition in the action as
\begin{eqnarray}
S_{GZ}&=& S_{YM} +\int d^4x \left[\bar\varphi_i^{a}\partial_\mu{\cal D}_\mu^{ab}
\varphi^{b}_i-\bar\omega_i^{a}\partial_\mu{\cal D}_\mu^{ab}
\omega^{b}_i\right.\nonumber\\
&-&\left.\gamma^2 g f^{abc} A^a_\mu(\varphi^{bc}_\mu
+\bar\varphi^{bc}_\mu) -4(N^2 -1)\gamma^4 
\right].
\end{eqnarray}
For the GZ action to be renormalizable, it is crucial to shift the field 
$\omega^a_i$, \cite{zwan2}
\begin{equation}
\omega^a_i(x)\rightarrow \omega^a_i +\int d^4y ({\cal M}^{-1})^{ab} (x, y) gf^{bkl}\partial_\mu
[{\cal D}_\mu^{ke} c^e (y)\varphi^l_i (y)],
\end{equation}
so that the complete GZ action becomes 
\begin{eqnarray}
S_{GZ}&=& S_{YM} +\int d^4x \left[\bar\varphi_i^{a}\partial_\mu{\cal D}_\mu^{ab}
\varphi^{b}_i-\bar\omega_i^{a}\partial_\mu{\cal D}_\mu^{ab}
\omega^{b}_i\right.\nonumber\\
&-&\left. gf^{abc}\partial_\mu\bar\omega^a_i{\cal D}_\mu^{bd}c^d\varphi^c_i 
-\gamma^2 g\left( f^{abc} A^a_\mu\varphi^{bc}_\mu\right.\right.\nonumber\\
&+&\left.\left. f^{abc} A^a_\mu\bar\varphi^{bc}_\mu +\frac{4}{g}(N^2 -1)\gamma^2 
\right)\right],\label{cgz}
\end{eqnarray}
which is multiplicative renormalizable.
\section{The nilpotent BRST transformations of GZ action}
The complete GZ action after localizing the nonlocal horizon term in D dimensional Euclidean 
space can be recast as
\begin{equation}
S_{GZ}=S_{exact}+S_\gamma \label{act}
\end{equation}
with $S_{exact}$, the BRST exact action and $S_\gamma$, the action for horizon term, defined
  as \cite{sor}
\begin{eqnarray}
S_{exact} &=& S_{YM} +\int d^4x \left[\bar\varphi_i^{a}\partial_\mu{\cal D}_\mu^{ab}
\varphi^{b}_i-\bar\omega_i^{a}\partial_\mu{\cal D}_\mu^{ab}
\omega^{b}_i\right.\nonumber\\
&-&\left. gf^{abc}\partial_\mu\bar\omega^a_i{\cal D}_\mu^{bd}c^d\varphi^c_i
\right],
\end{eqnarray}
\begin{eqnarray}
S_\gamma &=&-\gamma^2 g\int d^4x \left[f^{abc} A^a_\mu\varphi^{bc}_\mu +
f^{abc} A^a_\mu\bar\varphi^{bc}_\mu +\frac{4}{g}(N^2 -1)\gamma^2\right].\label{sgm}
\end{eqnarray}

The conventional BRST transformation for all the fields is given by
\begin{eqnarray}
\delta_b A_\mu^a &=& -{\cal D}_\mu^{ab}c^b\ \Lambda,\ \ \ \ \ \delta_b c^a =\frac
{1}{2}gf^{abc}c^bc^c
\ \Lambda, \nonumber\\
 \delta_b \bar c^a&=&B^a\ \Lambda,\ \ \ \ \ \ \ \ \ \ \delta_b B^a =0, \nonumber\\
 \delta_b\varphi_i^a &=& -\omega_i^a \ \Lambda,\ \ \ \ \ \ \ \ \ \delta_b\omega_i^a =0,
\nonumber\\
\delta_b\bar\omega_i^a &=&\bar \varphi_i^a \ \Lambda,\ \ \ \ \ \ \ \ \ \ \ \delta_b\varphi_i^a
 =0,\label{sym}
\end{eqnarray}
where $\Lambda$ is usual infinitesimal BRST parameter. 
But one can check that the BRST symmetry is broken softly for the GZ action \cite{zwan},
\begin{eqnarray}
\delta_b S_{GZ} &=& \delta_b (S_{exact} +S_\gamma )=\delta_b S_\gamma\nonumber\\
&=& \gamma^2 g\int d^4x f^{abc}\left ( A_\mu^a\omega_\mu^{bc}-({\cal D}_\mu^{am}c^m)
(\bar\varphi^{bc}_\mu +\varphi_\mu^{bc})\right ),
\end{eqnarray}
 the breaking is due to the presence of $\gamma$ dependent term, $S_\gamma$. 

To discuss the renormalizability of $S_{GZ}$, the breaking is treated as a composite operator
to be introduced into the action by means of a suitable set of external sources \cite{sor}. 
 Thus embedding the $S_\gamma$ in to a larger action with introducing 3 doublets of sources 
$(U_\mu^{ai}, M_\mu^{ai}), (V_\mu^{ai}, N_\mu^{ai}) $  and $(T_\mu^{ai}, R_\mu^{ai})$, as
\begin{eqnarray}
\Sigma_\gamma &=& \delta_b \int d^4 x \left (-U_\mu^{ai}{\cal D}_\mu^{ab}\varphi_i^b -
V_\mu^{ai}{\cal D}_\mu^{ab}\bar\omega_i^b -U_\mu^{ai}V_\mu^{ai} +gf^{abc}
T_\mu^{ai}{\cal D}^{bd}_\mu\bar\omega_i^c \right )\nonumber\\
&=&\int d^4 x \left ( -M^{ai}_\mu {\cal D}_\mu^{ab}\varphi^b_i 
-gf^{abc}U^{ai}_\mu{\cal D}_\mu^{bd}c^d\varphi^c_i +U^{ai}_\mu {\cal D}_\mu^{ab}
\omega^b_i\right. \nonumber\\
&-&\left. N^{ai}_\mu {\cal D}_\mu^{ab}\bar\omega^b_i -V^{ai}_\mu
{\cal D}_\mu^{ab}\bar\varphi^b_i +gf^{abc}V^{ai}_\mu{\cal D}_\mu^{bd}c^d
\bar\omega^c_i\right.\nonumber\\
&-&\left. M^{ai}_\mu V^{ai}_\mu +U^{ai}_\mu N^{ai}_\mu -gf^{abc}R^{ai}_\mu{\cal D}_\mu^
{bd}c^d\bar\omega^c_i +gf^{abc}T^{ai}_\mu{\cal D}_\mu^{bd}c^d\bar\varphi^c_i \right ),
\end{eqnarray}
whereas the sourcess involved $M_\mu^{ai}, V_\mu^{ai}, R_\mu^{ai}$ are commuting and 
$U_\mu^{ai}, N_\mu^{ai}, T_\mu^{ai}$ are fermionic in nature.
The above action is invariant under following BRST transformation
\begin{eqnarray}
\delta_b U^{ai}_\mu &=&M_\mu^{ai} \ \Lambda,\ \ \ \ \ \ \delta_b M^{ai}_\mu =0,\nonumber\\
\delta_b V^{ai}_\mu &=&-N_\mu^{ai}\ \Lambda,\ \ \ \ \ \ \delta_b N^{ai}_\mu =0,\nonumber\\
\delta_b T^{ai}_\mu &=&-R_\mu^{ai}\ \Lambda,\ \ \ \ \ \ \delta_b R^{ai}_\mu =0.\label{syma}
\end{eqnarray}
Therefore, the broken BRST has been restored at the cost of introducing new sources.
The different quantum numbers (to study the system properly) of fields and sources, 
involved in this theory, are discussed in Ref. \cite{sor}.
Still we do not want to change our original theory (\ref{sgm}).
Therefore, at the end,
we have to fix the sources equal to the following values:
\begin{eqnarray}
U^{ai}_\mu |_{phys} &=& N^{ai}_\mu |_{phys}=T^{ai}_\mu |_{phys}=0\nonumber\\
M^{ab}_{\mu\nu} |_{phys} &=& V^{ab}_{\mu\nu} |_{phys}=R^{ab}_{\mu\nu} |_{phys}=\gamma^2
\delta^{ab}\delta_{\mu\nu}.
\end{eqnarray}
It follows that $\Sigma_\gamma|_{phys}=S_\gamma $.  

The generating functional for the effective GZ action in Euclidean space is 
defined as
\begin{equation}
Z_{GZ}=\int [D\phi ] e^{-S_{GZ}},\label{zfun} 
\end{equation}
where $\phi$ is generic notation for all fields used in GZ action.

\section{FFBRST transformation in Euclidean space}
The properties of the usual BRST transformations do not depend on whether 
the parameter $\Lambda$  is (i) finite or infinitesimal, (ii) field dependent or not, as 
long 
as it is anticommuting and space-time independent. Keeping this in mind, Joglekar and 
Mandal introduced finite field dependent BRST transformation (FFBRST) \cite {sdj}, which has 
found many applications in gauge field theories \cite{sdj01, sdj4, sdj02, sdj03, sdj1, sdj3,
rb, sm, ssb, susk, subm}.
 These observations give us a freedom to 
generalize the  nilpotent BRST transformations in Eqs. (\ref{sym}) and (\ref{syma}) by
 the parameter, 
$\Lambda$ finite and field 
dependent without affecting its properties. To 
generalized the BRST transformations we start 
with making the  infinitesimal parameter field dependent by introducing a parameter $\kappa\ 
(0\leq \kappa\leq 1)$ and making all the fields, $\phi(x,\kappa)$, $\kappa$ dependent such 
that $\phi(x,\kappa =0)=\phi(x)$ and $\phi(x,\kappa 
=1)=\phi^\prime(x)$, the transformed field.

The usual infinitesimal BRST transformations, thus can be written generically as 
\begin{equation}
{d\phi(x,\kappa)}=\delta_b [\phi (x,\kappa ) ]\Theta^\prime [\phi (x,\kappa ) ]{d\kappa},
\label{diff}
\end{equation}
where the $\Theta^\prime [\phi (x,\kappa ) ]{d\kappa}$ is the infinitesimal but field 
dependent parameter.
The generalized BRST transformations with the finite field dependent parameter then can be 
constructed by integrating such infinitesimal transformations from $\kappa =0$ to $\kappa= 1$
, to obtain
\begin{equation}
\phi^\prime\equiv \phi (x,\kappa =1)=\phi(x,\kappa=0)+\delta_b[\phi(x) ]\Theta[\phi(x) ],
\label{kdep}
\end{equation}
where 
\begin{equation}
\Theta[\phi(x)]=\int_0^1 d\kappa^\prime\Theta^\prime [\phi(x,\kappa^\prime)],\label{thet}
\end{equation}
 is the finite field dependent parameter. Following this method, the modified BRST 
transformation, in Eq. (\ref{sym}), is generalized such that the parameter is finite and 
field dependent. 

Now we show that such offshell  nilpotent BRST transformations with finite field 
dependent parameter are symmetry  of the effective action in Eq. (\ref{cgz}). However, the 
path integral measure $[D\phi]$ in Eq. (\ref{zfun}) is not invariant under such 
transformations as the 
BRST parameter is finite. 

The Jacobian of the path integral measure in Euclidean space for such transformations can be 
evaluated for some 
particular choices of the finite field dependent parameter, $\Theta[\phi(x)]$, as
\begin{equation}
D\phi' =J(
\kappa) D\phi.
\end{equation}
The Jacobian , $J(\kappa )$ in the Euclidean space can be replaced (within the functional 
integral) as
\begin{equation}
J(\kappa )\rightarrow \exp[-S_1[\phi(x,\kappa) ]],
\end{equation}
 iff the following condition is satisfied \cite{sdj}
\begin{equation}
\int D\phi (x) \;  \left [ \frac{1}{J}\frac{dJ}{d\kappa}+\frac
{d S_1[\phi (x,\kappa )]}{d\kappa}\right ]\exp{[-(S_{GZ}+S_1)]}=0 \label{mcond}
\end{equation}
where $ S_1[\phi ]$ is local functional of fields.

The infinitesimal change in the $J(\kappa)$ can be written as
\begin{equation}
\frac{1}{J}\frac{dJ}{d\kappa}=-\int d^4x\left [\pm \delta_b \phi (x,\kappa )\frac{
\partial\Theta^\prime [\phi (x,\kappa )]}{\partial\phi (x,\kappa )}\right ],\label{jac}
\end{equation}
where $\pm$ sign refers to whether $\phi$ is a bosonic or a fermionic field.

Now, we generalize the BRST transformation given in Eqs. (\ref{sym}) and
(\ref{syma}) by making usual BRST parameter finite and field dependent as  
 \begin{eqnarray}
\delta_b A_\mu^a &=& -{\cal D}_\mu^{ab}c^b\ \Theta,\ \ \ \ \ \delta_b c^a =\frac
{1}{2}gf^{abc}c^bc^c
\ \Theta, \nonumber\\
 \delta_b \bar c^a&=&B^a\ \Theta,\ \ \ \ \ \ \ \ \ \ \delta_b B^a =0, \nonumber\\
 \delta_b\varphi_i^a &=& -\omega_i^a \ \Theta,\ \ \ \ \ \ \ \ \ \delta_b\omega_i^a =0,
\nonumber\\
\delta_b\bar\omega_i^a &=&\bar \varphi_i^a \ \Theta,\ \ \ \ \ \ \ \ \ \ \ \delta_b\varphi_i^a
 =0\nonumber\\
 \delta_b U^{ai}_\mu &=&M_\mu^{ai} \ \Theta,\ \ \ \ \ \ \delta_b M^{ai}_\mu =0,\nonumber\\
\delta_b V^{ai}_\mu &=&-N_\mu^{ai}\ \Theta,\ \ \ \ \ \ \delta_b N^{ai}_\mu =0,\nonumber\\
\delta_b T^{ai}_\mu &=&-R_\mu^{ai}\ \Theta,\ \ \ \ \ \ \delta_b R^{ai}_\mu =0,
\end{eqnarray}
where $\Theta$ is finite, field dependent, anticommuting and space-time independent parameter.
One can easily check that the above FFBRST transformation is also symmetry of the effective GZ action ($S_{GZ}$).
\section{A Mapping Between GZ theory and YM theory} 
In this section we establish the connection between the theories with GZ action and YM action 
by using finite field 
dependent BRST transformation. In particular, we show that the generating functional for 
GZ theory in path integral formulation is directly related to that of YM theory 
with proper choice of finite field dependent BRST transformation. The nontrivial Jacobian
of the path integral measure is responsible for such a connection. For this purpose
 we choose a
 finite field dependent parameter $\Theta$ obtainable from 
\begin{eqnarray}
\Theta^\prime &=&\int d^4x \left[\bar \omega^a_i\partial_\mu {\cal D}_\mu^{ab} 
\varphi_i^b
-U^a_\mu {\cal D}_\mu^{ab} \varphi_i^b -V^a_\mu {\cal D}_\mu^{ab}\bar \omega_i^b 
\right.\nonumber\\
&-&\left. U^{ai}_\mu V^{ai}_\mu +T^{ai}_\mu gf^{abc}{\cal D}_\mu^{bd}c^d\bar\omega^c_i\right].
\end{eqnarray}
using Eq. (\ref{thet}).
The infinitesimal change in Jacobian for above $\Theta^\prime$ using Eq. (\ref{jac}) is 
calculated as
\begin{eqnarray}
\frac{1}{J}\frac{dJ}{d\kappa}&=&-\int d^4x \left[-\bar \varphi^a_i\partial_\mu {\cal D}_\mu^
{ab}\varphi^b_i +\bar \omega^a_i\partial_\mu {\cal D}_\mu^{ab} 
\omega_i^b 
\right.\nonumber\\
&+&\left. gf^{abc}\partial_\mu\bar\omega_i^a{\cal D}_\mu^{bd}c^d\varphi^c_i +M^{ai}_\mu {\cal D}_\mu^{ab}\varphi^b_i -U^{ai}_\mu {\cal D}_
\mu^{ab}\omega^b_i 
\right.\nonumber\\
&+&\left. gf^{abc}U^{ai}_\mu{\cal D}_\mu^{bd}c^d\varphi^c_i +N^{ai}_\mu {\cal D}_\mu^{ab}\bar\omega^b_i +V^{ai}_\mu 
{\cal D}_\mu^{ab}\bar\varphi^b_i 
\right.\nonumber\\
& -&\left. gf^{abc}V^{ai}_\mu{\cal D}_\mu^{bd}c^d\bar\omega^c_i +M^{ai}_\mu V^{ai}_\mu -U^{ai}_\mu N^{ai}_\mu \right.\nonumber\\
& +&\left. gf^{abc}R^{ai}_\mu{\cal D}_\mu^
{bd}c^d\bar\omega^c_i -gf^{abc}T^{ai}_\mu{\cal D}_\mu^{bd}c^d\bar\varphi^c_i
\right].
\end{eqnarray}
Now the Jacobian for path integral measure 
in the generating functional (\ref{zfun}) can be replaced by $e^{-S_1}$ iff condition (\ref
{mcond}) is satisfied.
We consider an ansatz for $S_1$ as
\begin{eqnarray}
S_1&=&\int d^4x \left[\chi_1(\kappa)\bar \varphi^a_i\partial_\mu {\cal D}_\mu^
{ab}\varphi^b_i +\chi_2(\kappa)\bar \omega^a_i\partial_\mu {\cal D}_\mu^{ab} 
\omega_i^b 
\right.\nonumber\\
&+&\left. \chi_3(\kappa) gf^{abc}\partial_\mu\bar\omega_i^a{\cal D}_\mu^{bd}c^d\varphi^c_i 
+\chi_4(\kappa) M^{ai}_\mu{\cal D}_\mu^{ab}\varphi^b_i \right.\nonumber\\
&+&\left. \chi_5(\kappa) 
U^{ai}_\mu{\cal D}_
\mu^{ab}\omega^b_i +\chi_6(\kappa) gf^{abc}U^{ai}_\mu{\cal D}_\mu^{bd}c^d\varphi^c_i
\right.\nonumber\\
&+&\left.\chi_7(\kappa) N^{ai}_\mu{\cal D}_\mu^{ab}\bar\omega^b_i + 
\chi_8(\kappa) V^{ai}_\mu 
{\cal D}_\mu^{ab}\bar\varphi^b_i \right.\nonumber\\ 
&+& \left. \chi_9(\kappa) gf^{abc}V^{ai}_\mu{\cal D}_\mu^{bd}c^d
\bar\omega^c_i 
+ \chi_{10}(\kappa) M^{ai}_\mu V^{ai}_\mu \right.\nonumber\\
 &+&\left.\chi_{11}(\kappa) U^{ai}_\mu N^{ai}_\mu +
\chi_{12}(\kappa) gf^{abc}R^{ai}_\mu{\cal D}_\mu^
{bd}c^d\bar\omega^c_i
\right.\nonumber\\
& +&\left. \chi_{13}(\kappa) gf^{abc}T^{ai}_\mu{\cal D}_\mu^{bd}c^d\bar\varphi^c_i
\right].
\end{eqnarray}
where $\chi_j(\kappa) (j=1,2,.....,13)$ are arbitrary functions of $\kappa$ and satisfy 
following 
initial conditions
\begin{equation}
\chi_j(\kappa =0)=0.\label{inc}
\end{equation} 
The condition (\ref{mcond}) with the above $S_1$ leads to 
\begin{eqnarray}
\int &D\phi (x)&e^{-(S_{eff}+S_1)}\left[\bar \varphi^a_i\partial_\mu {\cal D}_\mu^
{ab}\varphi^b_i (\chi_1' +1) +\bar \omega^a_i\partial_\mu {\cal D}_\mu^{ab} 
\omega_i^b (\chi_2' -1) + gf^{abc}\partial_\mu\bar\omega_i^a{\cal D}_\mu^{bd}c^d\varphi^c_i
(\chi_3' -1)\right.\nonumber\\
&+&\left. M^{ai}_\mu {\cal D}_\mu^{ab}\varphi^b_i (\chi_4' -1)+ 
U^{ai}_\mu {\cal D}_
\mu^{ab}\omega^b_i(\chi_5' -1) + gf^{abc}U^{ai}_\mu{\cal D}_\mu^{bd}c^d\varphi^c_i (\chi_6' +1)
\right.\nonumber\\
&+&\left. N^{ai}_\mu {\cal D}_\mu^{ab}\bar\omega^b_i (\chi_7' -1) + 
V^{ai}_\mu 
{\cal D}_\mu^{ab}\bar\varphi^b_i (\chi_8' -1) + gf^{abc}V^{ai}_\mu{\cal D}_\mu^{bd}c^d
\bar\omega^c_i (\chi_9' +1)
\right.\nonumber\\
& +&\left.  M^{ai}_\mu V^{ai}_\mu (\chi_{10}' -1)+ U^{ai}_\mu N^{ai}_\mu (\chi_{11}' +1) +
 gf^{abc}R^{ai}_\mu{\cal D}_\mu^
{bd}c^d\bar\omega^c_i (\chi_{12}' -1)
\right.\nonumber\\
& +&\left.  gf^{abc}T^{ai}_\mu{\cal D}_\mu^{bd}c^d\bar\varphi^c_i (\chi_{13}' +1)
+ gf^{abc}\bar\varphi^a_i\partial_\mu{\cal D}_\mu^{bd}c^d\varphi^c_i\Theta (\chi_1 +
\chi_3)\right.\nonumber\\
&-&\left. \bar \varphi^a_i\partial_\mu {\cal D}_\mu^
{ab}\omega^b_i\Theta (\chi_1 +\chi_2 ) - gf^{abc}\bar\omega^a_i\partial_\mu{\cal D}_\mu^
{bd}c^d\omega^c_i\Theta (\chi_2-\chi_3)\right.\nonumber\\
& +&\left. gf^{abc}M^{ai}_\mu{\cal D}_\mu^{bd}c^d\varphi^c_i\Theta (\chi_4-\chi_5) 
- M_\mu^{ai}{\cal D}^{ab}_\mu\omega_i^b\Theta(\chi_4 +\chi_6)\right.\nonumber\\
& +& \left. gf^{abc}U^{ai}_\mu{\cal D}_\mu^{bd}c^d\omega^c_i\Theta (\chi_5 +\chi_6) 
- N_\mu^{ai}{\cal D}^{ab}_\mu\bar\varphi_i^b\Theta(\chi_7 -\chi_8)\right.\nonumber\\
&-&\left. gf^{abc}N^{ai}_\mu{\cal D}_\mu^{bd}c^d\bar\omega^c_i\Theta (\chi_7 +\chi_9) 
+  gf^{abc}V^{ai}_\mu{\cal D}_\mu^{bd}c^d\bar\varphi^c_i\Theta (\chi_8 +
\chi_9)\right.\nonumber\\
&+&\left. M^{ai}_\mu N^{ai}_\mu\Theta (\chi_{10}+\chi_{11})+
gf^{abc}R^{ai}_\mu{\cal D}_\mu^{bd}c^d\bar\varphi^c_i\Theta (\chi_{12} +\chi_{13})
\right] =0
\end{eqnarray}
where prime denotes the differentiation with respect to $\kappa$.
Equating the coefficient of terms $ \bar \varphi^a_i\partial_\mu {\cal D}_\mu^
{ab}\varphi^b_i,\ \bar \omega^a_i\partial_\mu {\cal D}_\mu^{ab} 
\omega_i^b,\  gf^{abc}\partial_\mu\bar\omega_i^a{\cal D}_\mu^{bd}c^d\varphi^c_i,\ 
M^{ai}_\mu {\cal D}_\mu^{ab}\varphi^b_i,$\\
$\ U^{ai}_\mu{\cal D}_
\mu^{ab}\omega^b_i, \ gf^{abc}U^{ai}_\mu{\cal D}_\mu^{bd}c^d\varphi^c_i, 
N^{ai}_\mu
{\cal D}_\mu^{ab}\bar\omega^b_i,\
 V^{ai}_\mu
{\cal D}_\mu^{ab}\bar\varphi^b_i $,\\ 
$ gf^{abc}V^{ai}_\mu{\cal D}_\mu^{bd}c^d
\bar\omega^c_i,\ M^{ai}_\mu V^{ai}_\mu,\  U^{ai}_\mu N^{ai}_\mu,\ gf^{abc}R^{ai}_\mu{\cal 
D}_\mu^{bd}c^d\bar\omega^c_i $ and $gf^{abc}T^{ai}_\mu{\cal D}_\mu^{bd}c^d\bar\varphi^c_i$
 from both sides
of above condition, 
we get following differential equations:
\begin{eqnarray}
\chi_1^\prime +1&=&0,\ \ \ \chi_2^\prime -1=0\nonumber\\
\chi_3^\prime -1&=&0,\ \ \ \chi_4^\prime -1=0\nonumber\\
\chi_5^\prime -1&=&0,\ \ \ \chi_6^\prime +1=0\nonumber\\
\chi_7^\prime -1&=&0,\ \ \ \chi_8^\prime -1=0\nonumber\\
\chi_9^\prime +1&=&0,\ \ \ \chi_{10}^\prime -1=0\nonumber\\
\chi_{11}^\prime +1&=&0,\ \ \ \chi_{12}^\prime -1=0\nonumber\\
\chi_{13}^\prime +1&=&0.\label{diff1}
\end{eqnarray}
The $\Theta$ dependent terms will cancel seperately and comparing the coefficients 
 of $\Theta$ dependent terms, we obtain 
\begin{eqnarray}
\chi_1 +\chi_2 &=&\chi_1 +\chi_3 =\chi_2-\chi_3 =\chi_4-\chi_5 = 0\nonumber\\
\chi_4 +\chi_6 &=&\chi_5 +\chi_6 =\chi_7-\chi_8 =\chi_7 +\chi_9 = 0\nonumber\\
\chi_8 +\chi_9 &=&\chi_{10} +\chi_{11} =\chi_{12}+\chi_{13}=0.\label{scond}
\end{eqnarray}
The particular solution of Eq. (\ref{diff1}) subjected to the condition (\ref{inc})
and Eq. (\ref{scond}) is
\begin{eqnarray}
\chi_1 &=& -\kappa,\ \ \chi_2 =\kappa, \ \ \chi_3 =\kappa, \ \ \chi_4 =\kappa \nonumber\\
\chi_5 &=& \kappa,\ \ \chi_6 =-\kappa, \ \ \chi_7 =\kappa, \ \ \chi_8 =\kappa \nonumber\\
\chi_9 &=& -\kappa,\ \ \chi_{10} =\kappa, \ \ \chi_{11} =-\kappa, \ \ \chi_{12} =\kappa \nonumber\\
\chi_{13} &=& -\kappa.
\end{eqnarray}
Therefore, the expression for $S_1$ in term of $\kappa$ is
\begin{eqnarray}
S_1&=&\int d^4x \left[-\kappa\ \bar \varphi^a_i\partial_\mu {\cal D}_\mu^
{ab}\varphi^b_i +\kappa\ \bar \omega^a_i\partial_\mu {\cal D}_\mu^{ab} 
\omega_i^b + \kappa\ gf^{abc}\partial_\mu\bar\omega_i^a{\cal D}_\mu^{bd}c^d\varphi^c_i
\right.\nonumber\\
&+&\left. \kappa\ M^{ai}_\mu{\cal D}_\mu^{ab}\varphi^b_i +\kappa\ 
U^{ai}_\mu {\cal D}_
\mu^{ab}\omega^b_i -\kappa\ gf^{abc}U^{ai}_\mu{\cal D}_\mu^{bd}c^d\varphi^c_i
\right.\nonumber\\
&+&\left. \kappa\ N^{ai}_\mu {\cal D}_\mu^{ab}\bar\omega^b_i + 
\kappa\ V^{ai}_\mu
{\cal D}_\mu^{ab}\bar\varphi^b_i -\kappa\ gf^{abc}V^{ai}_\mu{\cal D}_\mu^{bd}c^d
\bar\omega^c_i
\right.\nonumber\\
& +&\left. \kappa\ M^{ai}_\mu V^{ai}_\mu -\kappa\ U^{ai}_\mu N^{ai}_\mu +
\kappa\ gf^{abc}R^{ai}_\mu{\cal D}_\mu^
{bd}c^d\bar\omega^c_i
\right.\nonumber\\
& -&\left.\kappa\ gf^{abc}T^{ai}_\mu{\cal D}_\mu^{bd}c^d\bar\varphi^c_i
\right].
\end{eqnarray}
The transformed action is obtained by adding $S_1(\kappa=1)$ to $S_{GZ}$ as,
\begin{equation}
S_{GZ}+S_1=\int d^4x [\frac{1}{4}F_{\mu \nu }^aF_{\mu \nu}^a+B^a\partial_\mu A_\mu^a+i\bar 
c^a\partial_\mu {\cal D}_\mu^{ ab}c^b ].
\end{equation}
We left with the YM effective action in Landau gauge.
 \begin{equation}
 S_{GZ}+S_1=S_{YM}.
 \end{equation}
Note the new action is independent of horizon parameter $\gamma$, and hence horizon 
condition 
$\left(\frac{\partial \Gamma }{\partial \gamma^2}=0\right) $ 
 leads trivial relation for $S_{YM}$.
Thus using  FFBRST transformation we have mapped the generating functionals in 
Euclidean space as
\begin{equation}
Z_{GZ}\left(\int[ D\phi] e^{-S_{GZ}}\right)\stackrel{ FFBRST}{--\longrightarrow }Z_{YM}
\left(\int [D\phi] e^{-S_{YM}}\right),
\end{equation}
where $Z_{YM}$ is the generating functional for Yang-Mils action $S_{YM}$.
\section{Conclusion}
The GZ theory which is free from Gribov copies as the domain of integration is restricted 
to the first Gribov horizon, is not  invariant under usual BRST 
transformation due 
to the presence of the nonlocal horizon term. Hence the KO criterion for color 
confinement in a manifestly covariant gauge fails for GZ theory. A nilpotent BRST   
transformation which leaves GZ action invariant was developed recently and can be applied to 
KO analysis for color confinement. This nilpotent BRST symmetry is generalized by
allowing the transformation parameter finite and field dependent. This generalized BRST 
transformation is nilpotent and symmetry of the GZ effective action.
We have shown that this nilpotent
BRST with an appropriate choice of finite field dependent parameter can relate GZ theory 
with a correct horizon term 
to the YM theory in Euclidean space where horizon condition becomes a trivial one. Thus we 
have shown that 
theory free from Gribov copies (i.e. GZ 
theory with proper horizon term) can be related through a  nilpotent BRST 
transformation with a finite parameter to a theory with Gribov copies (i.e. YM 
theory in Euclidean space). 
The nontrivial Jacobian of such finite transformation is responsible for this important 
connection.

\acknowledgments
We thankfully acknowledge the financial support from the Department of Science and Technology 
(DST), Government of India, under the SERC project sanction grant No. SR/S2/HEP-29/2007.


\vspace{.2in}
\begin{thebibliography}{99}
\bibitem{fp} L. D. Faddeev and V. N. Popov, {\it{Phy. Lett.}} {\bf{B 25}}, 29 (1967).
\bibitem{gri} V. N. Gribov, {\it{Nucl. Phys.}} {\bf{B 139}}, 1 (1978).
\bibitem{zwan} D.Zwanziger, {\it{Nucl. Phys.}} {\bf{B 323}}, 513 (1989).
\bibitem{zwan2} D.Zwanziger, {\it{Nucl. Phys.}} {\bf{B 399}}, 477 (1993).
\bibitem{kon0} K. I. Kondo, {\it{Nucl. Phys.}} {\bf{B 129}}, 715 (2004).
\bibitem{zwan1} D.Zwanziger, {\it{Nucl. Phys.}} {\bf{B 378}}, 525 (1992).
\bibitem{ko} T. Kugo and I. Ojima, {\it{Suppl. Prog. Theor. Phys.}} {\bf{66}}, 1 (1979).
\bibitem{sore1} D. Dudal, J. A. Gracey, S. P. Sorella, N. Vandersickel and H. Verschelde 
{\it {Phys. Rev.}} {\bf{D 78}} 065047 (2008).
\bibitem{sor} D. Dudal, S. P. Sorella and N. Vandersickel  {\it{ Eur. Phys. J.}} {\bf 
C 68} 283 (2010).
\bibitem{dud} D. Dudal, S. P. Sorella, N. Vandersickel and H. Verchelde, {\it arXiv}:
{\bf 0904.0641} 
[hep-th].
\bibitem{fuj} K. Fujikawa, {\it{Nucl. Phys.}} {\bf{B 223}}, 218 (1983).
\bibitem{sor1} S. P. Sorella, {\it {Phys. Rev.}} {\bf{D 80}} 025013 (2009).
\bibitem{sdj} S. D. Joglekar and B. P. Mandal, {\it {Phys. Rev.}} {\bf{D 51}} 1919 (1995).
\bibitem{kon1} K. I. Kondo, {\it{ arXiv}}: {\bf{0909.4866}}[hep-th].
\bibitem{sdj01} S. D. Joglekar and A. Misra, Int. J. Mod. Phys. {\bf{A 15}}, 1453 (2000).
\bibitem{sdj4}R. S. Bandhu and S. D. Joglekar, {\it{J. Phys.}} {\bf{A 31}}, 4217 (1998).
\bibitem{sdj02} S. D. Joglekar and A. Misra, {\it{ Mod. Phys. Lett.}} {\bf{A 14}}, 2083 
(1999).
\bibitem{sdj03} S. D. Joglekar and A. Misra, {\it{Mod. Phys. Lett.}} {\bf{A 15}}, 541 (2000).
\bibitem{sdj1} S. D. Joglekar and A. Misra, {\it{J. Math. Phys}} {\bf{41}}, 1755,(2000).
\bibitem{sdj3} S. D. Joglekar and B. P. Mandal, {\it {Phys. Rev.}} {\bf{D 55}}, 5038 (1997).
\bibitem{rb} R. Banerjee and B. P. Mandal Phys. Lett. {\bf{B27}} 488 (2000).
\bibitem{sm} S. K. Rai and B. P. Mandal,  Euro. Phys J {\bf C63}, 323,  (2009).
\bibitem{ssb} B. P. Mandal, S. K. Rai and S. Upadhyay, {\it Euro. Phys. Lett.} {\bf 92}, 
21001 (2010).
\bibitem{susk} S. Upadhyay, S. K. Rai and B. P. Mandal, {\it{ arXiv}}: {\bf
{1002.1373}}[hep-th] {\it to appear in J. Math. Phys}.
\bibitem{subm} S. Upadhyay  and B. P. Mandal, {\it Mod. Phys. Lett.  }{\bf A 25}, 3347 
(2010).


\end{thebibliography}
\end{document}